# Field Ionization of Cold Atoms near the Wall of a Single Carbon Nanotube

Anne Goodsell[1], Trygve Ristroph[1], J. A. Golovchenko[1], and Lene Vestergaard Hau[1]

*[1]Department of Physics, Harvard University, Cambridge MA 02138, USA*

We observe the capture and field ionization of individual atoms near the side wall of a single suspended nanotube. Extremely large cross sections for ionization from an atomic beam are observed at modest voltages due to the nanotube's small radius and extended length. The effects of the field strength on both the atomic capture and the ionization process are clearly distinguished in the data, as are prompt and delayed ionizations related to the locations at which they occur. Efficient and sensitive neutral atom detectors can be based on the nanotube capture and wall ionization processes.

The ability to create extremely high electric fields near charged carbon nanotubes has stimulated the development of electron field emission sources [1,2], with potential application in low power video displays, and of sensitive gas sensors whose ionization currents depend on the pressure and species present [3-5]. The quantitative behavior of many of these devices depends on the fact that the large fields required are obtained at the tip of a nanotube or, more commonly, at the tips of an inhomogeneous forest of nanowires. Variability of nanotube lengths, density, and tip geometry complicates the interpretation of the resulting device behavior. Also, the effective active area of the device is often a small fraction of that occupied by nanotubes.

This situation is dramatically modified when the side wall of a charged, suspended nanotube is used as the source of the high electric field rather than its tip. Here we demonstrate with a single nanotube that field ionization of ground state rubidium atoms can be effected with very high efficiency at modest voltages. Extremely large cross sections for ionization from an atomic beam are observed due to the nanotube's small radius, extended length, and the peculiar dynamics of a polarizable neutral atom near an extended charged wire. Our studies are carried out with a pulsed source of ultra cold atoms, which increases the signal-to-noise ratio of the measurements by several orders of magnitude over what can be obtained with thermal atoms. This allows studies of the dynamics of both the capture and ionization processes for individual atoms interacting with a single charged nanotube.

The setup demonstrated here can be applied directly as a compact, chip-integrated neutral atom detector with single atom sensitivity even for ground state atoms, and with high spatial and temporal resolution, and species selectivity. This is a long sought goal and has applications for sensitive gas detection, the development of compact, cold-atom based interferometers, and atom counting and quantum correlation measurements in cold atomic gases [6-10]. The system can also be used to manipulate quantum degenerate gases into extreme conditions heretofore unexplored.

Linearly polarizable atoms from an external beam are attracted by a $1/r^2$ potential to a uniformly charged, thin, extended wire. This is the same radial dependence as the repulsive centrifugal barrier, and classical equations of motion predict that there are two categories of trajectories [11]. Atoms that approach the wire with angular momentum L less than a critical value $L_c$ will be captured into trajectories that spiral towards the wire with no possibility of escape. They will inevitably impinge on the surface of the wire. Their orbits are described in cylindrical coordinates by

$$r(\theta) = \frac{bk}{\sinh(k\theta)},$$

where $k^2 = L_c^2/L^2 - 1$, and b is the impact parameter. Atoms with angular momentum greater than $L_c$ initially spiral in towards the wire but ultimately escape to infinity without being captured. $L_c$ is proportional to the voltage, $V_{NT}$, on the nanotube: $L_c = V_{NT}\sqrt{\alpha m}/\ln(R/r_{NT})$, where $r_{NT}$ and $R$ are the radius of the nanotube and an effective outer grounding cylinder, $\alpha$ is the atomic DC polarizability, and $m$ is the atomic mass. (We use R=1 cm but the results are rather insensitive to this value as it appears only within a logarithm.) The tube therefore acts as a "black hole", capturing those incoming atoms that have impact parameter below a critical value, $b_c = L_c/mv$, that scales linearly with the applied voltage $V_{NT}$ and inversely with the launch velocity v. With the nanotube at 300 V, the critical angular momentum is ~5000 $\hbar$ for rubidium atoms, and for an atom velocity of 5.3 m/s, $b_c$ ~750 nm, more than 2 orders of magnitude larger than a typical nanotube diameter. An atom with an impact parameter just below this critical value at first orbits the nanostructure and then enters a phase of dramatic collapse towards it. The kinetic energy of the accelerating atom increases by 4 orders of magnitude and the orbit time decreases to picoseconds. For high enough positive charging voltage on the nanotube, the atom's valence electron will tunnel into the nanotube, transforming the atom to an ion that is then ejected with high kinetic energy from the vicinity of the nanotube by Coulombic repulsion. The capture and ionization of single atoms is observed by detecting the ejected ion with a channel-electron multiplier (CEM) [12].

In the experiments, we use a suspended, single-walled carbon nanotube, with a diameter of a few nanometers and a length of 10 μm. The nanotube was grown by chemical vapor deposition [1,13] across an ion-milled gap in a free-standing, insulating silicon nitride membrane on a silicon chip, and then electrically contacted at both ends with shadow-masked evaporated metallic electrodes (see supplementary information [14]). Voltages in excess of 300 V can be applied to the nanotube without inducing electrical breakdown to the grounded silicon substrate below the contacts or damaging the nanotube. Figure 1(a) shows a scanning electron microscope (SEM) image of a free-standing nanotube suspended on its chip, and Fig. 1(b) presents an expanded schematic illustrating how a cold atom interacts with the tube.

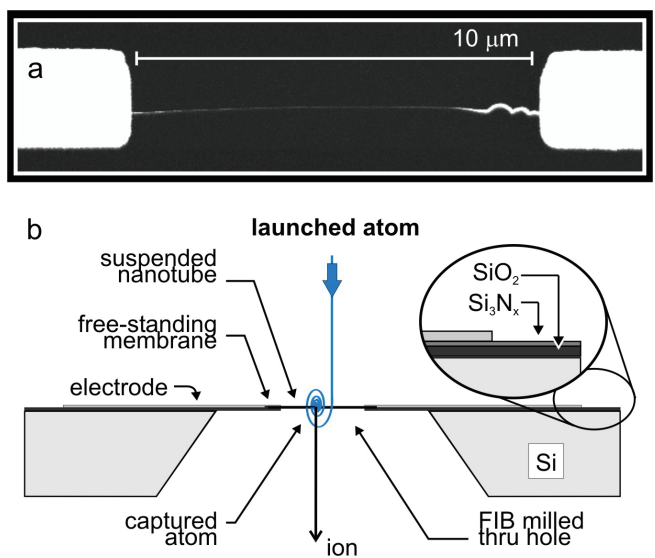

FIG. 1 (color online). Suspended nanotube for atom capture. (a) SEM image of the top surface of the sample, showing the nanotube suspended between rigid membrane arms of silicon oxide and silicon nitride. This sample yielded the data presented here. (b) Schematic cross-section of the sample chip with trajectory of an atom that is captured and ionized.

Rubidium atoms are cooled in a magneto optical trap (MOT) to 200 μK, corresponding to an rms velocity of 0.2 m/s, and launched once every 2 sec towards the voltage-biased nanotube at a drift velocity of 5.3 m/s [15]. Each cold-atom pulse contains $10^6$ atoms and has a simple, initial transverse diameter of 0.55 mm and a more complex longitudinal profile that will be described below. Figure 2 is a schematic of the apparatus. Shown at the bottom is a fluorescence image of rubidium atoms after cooling in the MOT. These atoms are optically launched upward towards the carbon nanotube sample, and the absorption profile of each atom pulse is monitored with a horizontally propagating laser beam positioned just below the sample. A shielded CEM sits above the nanotube and counts single ions emanating from it.

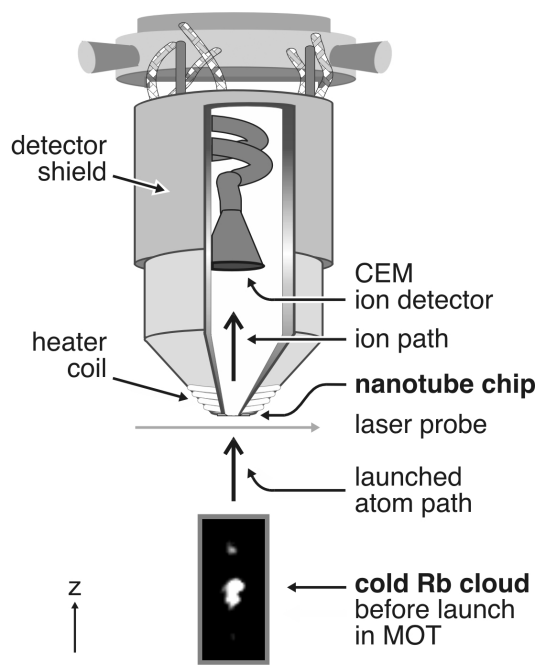

FIG. 2. Cold-atom nanotube apparatus. Rubidium atoms are loaded and laser-cooled in a stationary MOT. The 550-μm diameter atom cloud (fluorescence image shown at bottom) is launched and the pulse travels 22 mm to the nanotube. Ions are detected 25 mm above the sample by the CEM (kept at -2.3 kV). A heater coil controls the sample temperature, and a shield around the ion detector minimizes background counts.

Figure 3 shows the time distribution of detected ions measured from the launch. The figure also shows the corresponding optical density profile of the incident atomic pulse as it passes the probe laser beam. The pulse contains a main lobe, approximately 300 μsec in width, and a secondary structure leading it by 700 μsec. The ion signal has a slightly narrower time spread consistent with the finite (210 μm) size of the probe laser beam. All nanotube capture results presented below have been normalized to the incident atomic beam intensity deduced from optical absorption measurements.

Figure 4(a) shows the average number of ions detected per launched atomic pulse as a function of nanotube voltage from 0 to 300 V. (For each pulse, ions are counted in the time interval $3.0 \leq t \leq 5.6$ msec as determined from Fig. 3.) The detected ion signal is proportional to the voltage in the region above 150 V, which is a direct manifestation of the dynamics of the capture process and corresponds to the linear increase of the critical impact parameter with voltage as discussed above [11,16,17]. At the highest voltages we detect 1/3 ion per launch. From the optical density measurements and the measured transverse atomic cloud size at the sample position, we estimate that 5-10 atoms are captured per launch. The result is a detection efficiency of 5% for atoms incident within the critical impact parameter, which is consistent with our CEM detector efficiency and solid angle considerations.

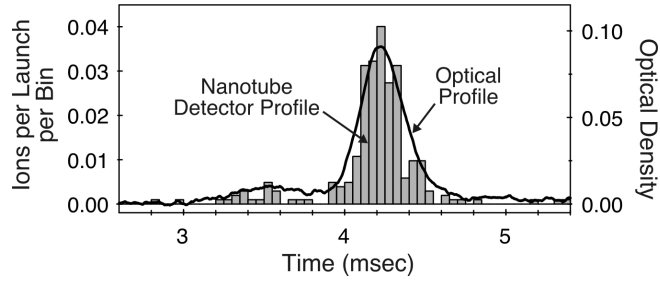

FIG. 3. Optical and nanotube detection of launched cold atoms. The ion signal is measured with the CEM after atoms are ionized by the nanotube kept at high voltage (>200V). The data show the temporal profile of an atom cloud as it arrives at the sample, and are compared to the optical density (OD) profile measured with the laser beam. The histogram has a bin size of 50 μsec and a total of 259 single ions. For direct comparison, the OD signal is offset by 0.33 msec, the time-of-flight for atoms between the laser probe and the sample.

Interestingly, the ion signal in Fig. 4(a) increases abruptly from zero over a 25 V range starting at 125 V. This can be explained by considering the process of electric-field stimulated electron tunneling close to the nanotube. The rubidium valence electron tunnels across the potential barrier consisting of the combined electrostatic potential from the $Rb^+$ core and the charged nanotube, thereby creating an ion. The tunneling rate as determined from a WKB approximation [18] increases exponentially with the strength of the electric field generated by the charged tube at the location of the atom. The electric field varies inversely with the atom's distance from the tube's center, and the tunneling rate rapidly rises to $10^{12}$/s as the field reaches 3 V/nm. For each voltage, we calculate the probability that captured atoms will ionize as they spiral towards the nanotube. With a tube radius of 3.3 nm we find that the ionization begins when the voltage reaches 125 V, and for voltages above 150 V the field is so large that a captured atom will ionize with 100% probability. The value of the threshold voltage is a sensitive measure of the nanotube radius and the ionization energy of the atom. As shown in Fig. 4(a), there is excellent agreement between experimental data and this model (solid curve).

We also measured the arrival time of single ions at the CEM after the atom pulse is launched from the MOT. Arrival events are recorded (with a resolution of 60 nsec) in a time window spanning 50 msec both before and after the launch time. In addition to monitoring the (negligible) background count rate (3.0 ions/sec at 300 V), this measurement reveals that, for voltages near the threshold for ionization, many ions arrive at the detector with significant time delays. This is easily seen in Figs. 4(b) and 4(c), which compare the distribution of arrival times for voltages in the 125-175 V range to that in the 250-300 V regime. For the lowest voltages, the ionization occurs within 0.4 nm from the nanotube surface and the ion becomes trapped by its induced image potential (the relevant image potential was derived in Ref. [19]). Such ions are temporarily bound near the surface, initially orbiting the nanotube with tens of picoseconds orbit times, and they escape only after a delay time that is determined by their ability to overcome the binding energy, either thermally or via surface defect interactions.

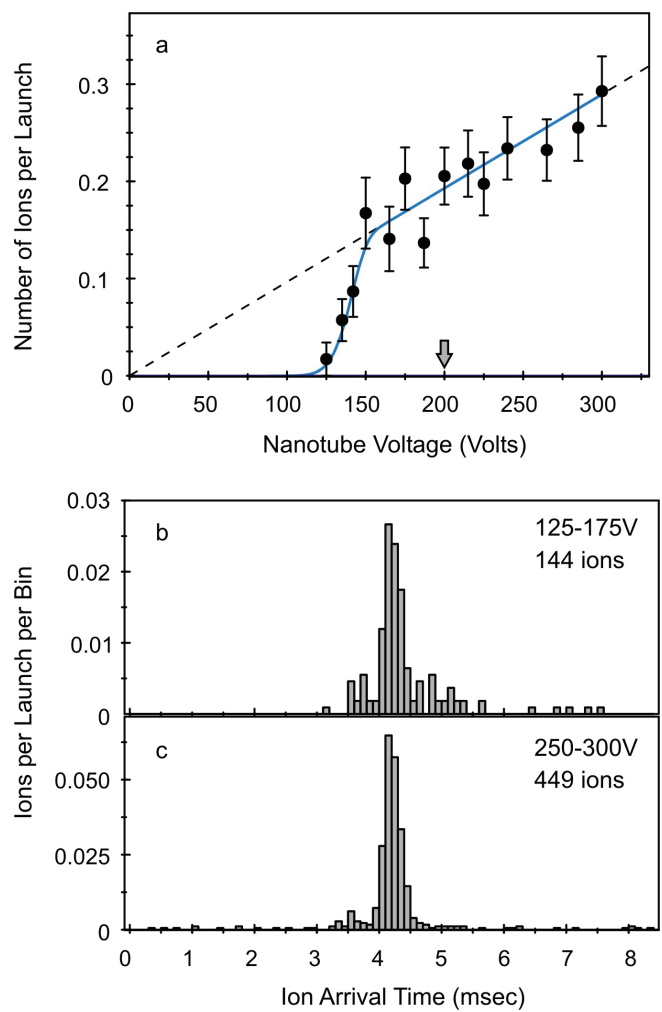

FIG. 4 (color online). Number of ions vs. nanotube voltage, and interaction dynamics at the nanoscale between cold atoms and a nanotube. (a) Experimental measurements (points) and theoretical model (solid curve) of the number of ions detected from launched atom pulses with velocity 5.3 m/s and peak OD = 10%. The model has a nanotube radius of 3.3 nm. The threshold behavior (125-150 V) shows that a minimum electric field is needed for efficient electron tunneling, as described in the text. The arrow marks the voltage above which ion arrivals are prompt indicating that electron tunneling takes place outside the image potential barrier. The dashed line reflects the linear voltage dependence of the critical impact parameter for atom capture and corresponds to what we would measure if atoms were ionized with 100 % probability at all voltages. It agrees with the data in the high-voltage regime. Error bars (s.d.) are determined from counting statistics. (b), (c) Measured ion arrival times relative to the atom launch at t = 0 msec, with 100 μsec bin size. Background counts are negligible as the time frame from 0-3 msec shows.

Consider the thermal process. The radial, image-potential surface trap frequency is ~$10^{13}$ Hz, and for an ion in thermal equilibrium with the nanotube (kept at the substrate temperature of 100°C) a Boltzmann distribution of energies is expected. The escape rate of a bound ion from the trapping potential is determined by the product of the trap frequency and the fraction of ions in the Boltzmann tail above the trap barrier. The barrier is dependent on the nanotube voltage and the minimum of the trap potential. This picture accounts for the observed millisecond delay times in the voltage range shown in Fig. 4(b) if, in the model, we use a barrier height of 0.8 eV above the minimum. A similar value was found in Ref. [20] for the binding energy of a (charged) rubidium atom to an uncharged nanotube.

To complete the picture, for voltages above 200 V [to the right of the arrow in Fig. 4(a)] the electric fields are large enough to cause ionization of the incident neutral atoms before they enter the image-potential barrier, and these ions are therefore ejected promptly from the tube region in agreement with the prompt signals of Fig. 4(c).

Thus we have observed the capture and ionization of individual atoms at the wall of a single suspended nanotube. The cross sections for side wall ionization that we measure are five (three) orders of magnitude larger than the calculated cross sections for tip ionization of cold (thermal) atoms for electric fields kept below the field evaporation limit [21,22]. The system operated as an atom detector has a time resolution for single atom detection in the nanosecond regime (for voltages above 200 V), and a spatial resolution that is currently determined by the capture cross-section of the nanotube. The threshold voltage is a sensitive probe of the species being ionized, and with a modified geometry and ion optics, the detector's efficiency for atom counting should approach 100%. With a position-sensitive microchannel plate (instead of the CEM) and ion optics, spatial resolution at the single nanometer level can be obtained along the length of the tube. Combined with the large atom capture range perpendicular to the tube, this allows direct measurement of fringes of interfering matter waves in a chip-based atom interferometer.

Quantization of angular momentum is predicted to result in step increases of the atom capture cross-section for increasing tube voltage (60 mV/$\hbar$ for rubidium) [17]. With improved measurement statistics, these steps can be revealed. Further control over cold atom dynamics can be effected by addition of an ac potential to the nanotube and leads to stable atom trapping around the tube for extended timescales [11]. Atoms will be captured from a distance of several microns, and the radius of the trapped orbit can then be lowered adiabatically to a few tens of nanometers. For multiple, trapped atoms, the induced electric dipole moments will lead to strong dipole-dipole coupling and potentially to novel, highly correlated states [23,24] that can be probed directly with the nanotube detector.

We acknowledge contributions from M. Burns, F. Gong, J. Welch, J. MacArthur, S. Garaj, and C.N.T. Lucy.  This work was supported by the Air Force Office of Scientific Research and the National Science Foundation. Some fabrication was carried out at Harvard University's Center for Nanoscale Systems, with assistance of D. Bell and Y. Lu.

## Supplementary Information for
## Field Ionization of Cold Atoms Near the Wall of a Single Carbon Nanotube

Anne Goodsell, Trygve Ristroph, J. A. Golovchenko, and Lene Vestergaard Hau

**Preparation of samples with suspended nanotubes.**

Single-walled carbon nanotubes, suspended across 10-μm gaps, are grown on silicon chips in a CVD oven. Each sample is 3mm x 3mm and cut from a 525-μm thick silicon wafer (p-type with resistivity of 10-20 Ωcm at 300K) with an overlayer of $SiO_2$ and low-stress $Si_3N_x$ on both sides. After using photolithography and a reactive ion etch to selectively remove the bottom layer of $Si_3N_x$, we apply an anisotropic silicon etch that creates a free-standing 70 x 70 $\mu m^2$ membrane consisting of 2 μm of $SiO_2$ and 200 nm of $Si_3N_x$. A through-hole in the membrane, with a shape as shown in Supp. Fig. 1b, is milled with a focused ion beam (FIB). With a shadow mask we then e-beam evaporate catalyst pads, consisting of 1 nm Fe on top of 10 nm $Al_2O_3$, [25] at the ends of the long membrane arms (Supp. Fig. 1a).

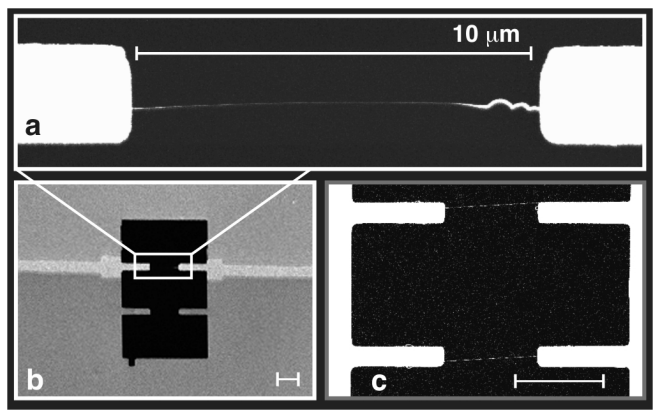

**Supplementary Figure 1. Suspended nanotubes for atom capture.** The SEM images show the top surfaces of the nanotube samples. The chip structures are shown with suspended nanotubes and Ti/Mo electrodes that are insulated by SiO/SiN from the silicon substrates. The scale bar in all images is 10 μm. The sample shown in (a, b) yielded the data presented in the main text. Figure **a,** a single tube is suspended between rigid SiO/SiN arms, with metal evaporated on the arms and at the ends of the tube; **b,** thru-hole (dark) milled in the rigid, free-standing membrane, bridged by the tube shown above, and with narrow electrodes (pale) deposited after tube growth; **c,** sample with two parallel, suspended tubes shown before electrodes are evaporated (we use high contrast here to show the nanotube locations).

Nanotubes are grown – from the catalyst pads and between the arms – with chemical vapor deposition in an oven heated to 850C and with methane as the carbon source [1,13]. Often we observe several tubes on a sample as seen in Supp. Fig. 1c. We finally evaporate electrodes with 10 nm of Mo on top of 50nm of Ti. (No degrading of the electrodes is observed even after extended periods of exposure to rubidium). A shadow-mask stencil is milled with the FIB, mounted in a translatable holder with ~20 μm clearance over the silicon chip, and then aligned under an optical microscope before electrode deposition in the e-beam evaporator [26]. While a wet-lithography process can cause contamination or damage for long suspended tubes, the post-growth electrode patterning we use leaves the tubes unharmed. The Ti/Mo strips cover the top surface of the membrane arms, creating well-defined 2-µm-wide electrodes, and wet the ends of the suspended tube for improved contact [27] (Supp. Fig. 1a, b). The strips are contacted to 2-mm-sized macroscopic pads (outside the field of view of the images), and a high voltage is applied with clip connections to the pads. The electrode design was optimized with numerical modelling based on a finite-element analysis with COMSOL Multiphysics® (a commercial software package by COMSOL AB). In the analysis we include the full three-dimensional structure of the chip and confirm that electric fields are determined entirely by the charge density on the nanotube and that shielding effects from the 2 -µm wide electrodes are negligible.

When a sample is placed in the vacuum chamber ($10^{-9}$ torr), the silicon is grounded and the electrodes are charged to positive voltage to generate the atom-capturing field. The sample is mounted upside-down as shown in Fig. 2 of the main text. The cone-shaped sample holder, with the chip mounted at the apex, conforms to the geometry of our laser beams, and x-y-z translation stages allow continuous scanning of the sample in three dimensions. Using the wrapped-wire heater, the sample temperature is slightly elevated (100°C) to prevent the accumulation of rubidium on the nanotube chip.

**MOT loading and launching.**

We operate a 2D magneto-optical trap (MOT) in pulsed mode to cool, confine, and launch clouds of $^{85}$Rb atoms [15]. We use four trapping laser beams and elongated current coils that generate magnetic gradients in the two horizontal directions. The MOT is vapor-loaded from an evaporable getter source for 1.25 sec, resulting in a trapped cloud with 1 million atoms and a temperature of 200 µK. To launch an atom cloud, we abruptly change the relative detuning $\delta$ between two pairs of laser beams and create a moving MOT. The launch velocity $v$ is proportional to $\delta$, with $v/\delta$=(1 m/s)/(1.81Mhz). We chose a nominal launch velocity of 5.0 m/sec ($\delta$= 9.06 Mhz), and the actual atom velocity was determined by time-of-flight measurements to be $v$ =5.3 m/sec. (This difference is due to slight errors in optical beam alignment for the MOT). Importantly, the launch velocity was chosen to be much larger than the rms velocity of 20 cm/s of the atom cloud, and all atoms are therefore launched toward the tube with nearly the same velocity. Atoms arrive at the tube 4.2 ms after being launched. For each launch, the time-dependent optical density of the atom cloud is measured by a weak, resonant laser probe beam (size of 140 µm x 210 µm and power of 100nW) placed just below the nanotube chip, and we measure a peak density of $1.2 \times 10^9$ atoms/cm$^3$ corresponding to a peak flux of $6.4 \times 10^{11}$ atoms/cm$^2$/sec. The size of the launched atom cloud in the horizontal plane at the sample position is 1100 µm by 800 µm (FWHM), which is measured with the nanotube by scanning it across the atomic beam profile.

**Ion detection and time stamping.**

For ion detection, we use a single-channel electron multiplier (Channeltron® CEM, manufacturer Burle/Photonis) run in discrete pulse-counting mode. The funnel-shaped detector (main text Fig. 2) is charged to –2300V; the rim has a transverse diameter of 9.9 mm and is positioned 25 mm from the nanotube. Individual ions incident on the CEM generate discrete charge pulses that are input to a charge-sensitive amplifier (Amptek A121 discriminator/preamp that requires a minimum pulse-to-pulse separation of 60 nsec). The timing of the output signals is digitally recorded with an accuracy of 14 nsec. For each launch we generate a list of time-stamps corresponding to every ion arriving during a 100 msec time window, including a 50 msec time interval both before and after atom launch.

To produce Fig. 4a (in main text), we distinguish those ions which correspond to captured cold atoms from background ions by counting only the ions that arrive within a 2.6 msec time window ($3.0 \leq t \leq 5.6$ msec) determined from the histograms of ion arrival times. Counts from ionization of the background rubidium vapor are accumulated in the time window before the launch ($-50 \leq t \leq 0$ msec). The background rate is ~ 0.5-3.0 ions/sec, increasing with voltage, so the contribution is negligible during the short time window of the cold-atom pulse and background subtraction is unnecessary.

**Interaction potentials between a carbon nanotube and neutral or singly charged rubidium.**

In Supp. Fig. 2 we show the attractive $1/r^2$ potential [11] that governs the motion of a neutral rubidium atom in the electric field from a charged carbon nanotube. In Supp. Fig. 3 we show the interaction potential between a rubidium ion and a charged carbon nanotube. The potential is a combination of the image potential induced by the ion [19] and the repulsive potential from the positively charged tube. When the atom ionizes within the potential barrier at 0.4 nm, the ion is trapped near the surface as described in the main text.

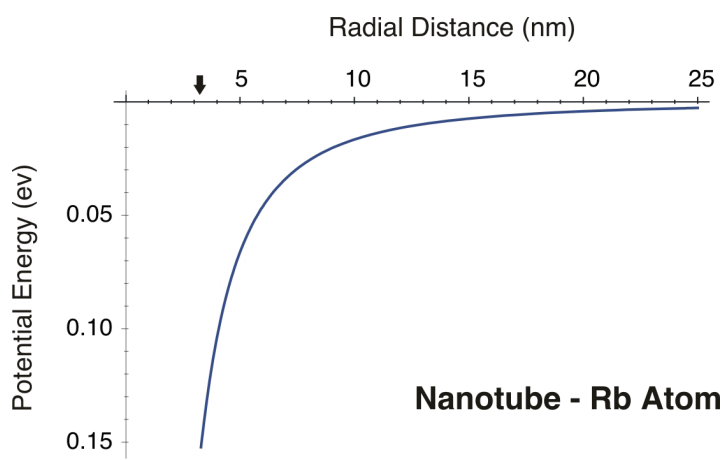

**Supplementary Figure 2. Attractive $1/r^2$ potential for a neutral rubidium atom around a carbon nanotube charged to 150V.** The arrow indicates the surface of the 3.3 nm radius nanotube used in the experiments. The polarizability of Rb is 47 $\text{Å}^3$.

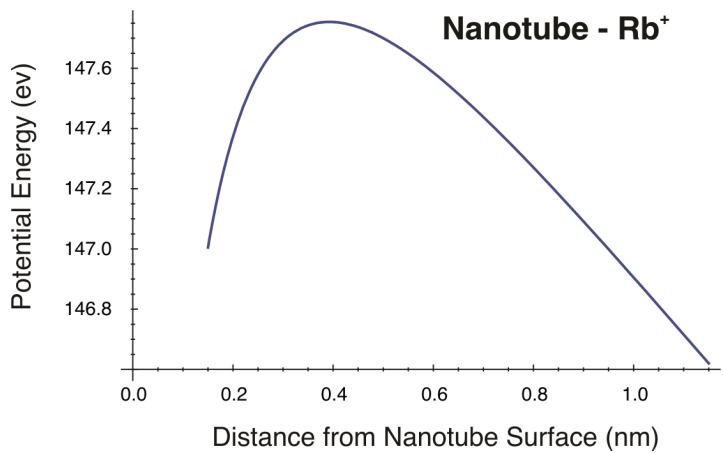

**Supplementary Figure 3. Potential for a Rb⁺ ion in proximity to a charged carbon nanotube wall**. This is a combination of the ion's induced image potential and the repulsive potential from a nanotube positively charged to 150V. The nanotube radius is 3.3 nm. From the measurements of nanoscale interaction dynamics (main text Fig. 4b, c) we estimate repulsive interactions (not shown) to be dominant for distances below 1.5 Å. An ion, created at this tube voltage by field ionization of a captured rubidium atom, is trapped in the resulting potential minimum for roughly a millisecond.

**Supplementary References**